%% file: directed_communities.tex
\documentclass{ase}  
\usepackage{amsmath,amsfonts,amssymb}

\usepackage{graphicx}
\usepackage{epstopdf}
\usepackage{booktabs}
\usepackage{tabularx}
\usepackage{float}

\usepackage{subfigure}
\usepackage{tabularx}
\usepackage{multirow}

\usepackage{enumitem}

\usepackage{braket}

\usepackage{tikz}
\usetikzlibrary{arrows,positioning,backgrounds}

\usepackage{wrapfig}

\usepackage{capt-of}
\hypersetup{colorlinks,urlcolor=blue,citecolor=blue,linkcolor=blue}

\begin{document}

\begin{center}

\huge Using Triangles to Improve Community Detection \\[6pt] in Directed Networks

\normalsize

% Author block

%At the end of each author, use \\ \columbreak to move to the next column as shown
\begin{multicols}{3} %you may adjust this as needed
{Christine Klymko}{\\}{Lawrence Livermore National Laboratory\\Email: klymko1@llnl.gov \\ \columnbreak}
{David Gleich}{\\}{Purdue University\\Email: dgleich@purdue.edu \\ \columnbreak}
{Tamara G. Kolda}{\\}{Sandia National Laboratories\\Email: tgkolda@sandia.gov}\\

\end{multicols}
\end{center}

\begin{multicols*}{2}

%\pagenumbering{arabic}
%\setcounter{page}{1}%Leave this line commented out.

\begin{abstract}
  In a graph, a community may be loosely defined as a group of
  nodes that are more closely connected to one another than to the
  rest of the graph. While there are a variety of metrics that can be used
  to specify the quality of a given community, one common theme is that
  flows tend to stay within communities. Hence, we expect
  cycles to play an important role in community detection. For
  undirected graphs, the importance of triangles -- an undirected
	3-cycle -- has been known for a long time and can be used to improve
  community detection. In directed graphs, the situation is more
  nuanced. The smallest cycle is simply two nodes with a reciprocal
  connection, and using information about reciprocation has proven to
  improve community detection. Our new idea is based on the four types
	of directed triangles that contain cycles. 
To identify communities in directed 
networks, then, we propose an undirected edge-weighting scheme based on the type of the directed triangles 
in which edges are involved. We also propose a new metric on quality of the communities 
that is based on the number of 3-cycles that are split across communities. To demonstrate the impact of our new weighting, we use the standard METIS
graph partitioning tool to determine communities and 
show experimentally that the resulting communities result in fewer 3-cycles being cut. The magnitude of the effect varies between a 
10 and 50\% reduction, and we also find evidence that this weighting scheme improves a task where plausible ground-truth communities are known.

{\bf keywords:} community detection, directed networks, triangles, reciprocity, 3-cycles
\end{abstract}

%% INTRODUCTION
\input{Introduction}

\input{Background}

\input{Metric}

\input{Methods}

\input{Evaluation}

%\section{Significance and Impact.}

%\subsection{Community quality based on 3-cycles}

\section{Conclusions.}
\input{Conclusions}

\bibliographystyle{IEEE}
\bibliography{directed_communities,gleich}
\end{multicols*}

\end{document}

%% file: Introduction.tex
\section{Introduction}

Many different systems can be viewed as complex networks made up of objects (nodes) and 
connections between them (links or edges).   Over the past several decades the study of such
networks has become important in many disciplines \cite{BoLaMoChHw06,BrEr05,Es11,Ne10},
and a recurrent research theme is finding the communities or modules within 
these networks. These communities reveal important structures hidden within
the network data.

Thus far, 
the majority of work in community detection has focused on undirected 
networks (see the survey~\cite{Fo10}), although, recently, more research has 
focused on directed networks (see the survey~\cite{MaVa13}).  
In the most common setting, a community is a group of nodes that 
are more closely connected to each other than to the rest of the network.  
Community assignment methods use only topological features of the network unless 
additional information about the components of the network is known; and
thus the connectivity between nodes is often used alone to define metrics 
measuring the ``quality'' of the assigned groups.  Common quality
metrics measure (i) the density of links
within a group (modularity) \cite{Ne03}, (ii) the number of cut edges relative
to the group size (conductance), (iii) the stationary distribution 
of a random walk within the network (LinkRank) \cite{KiSoJe10}, (iv)
and the probability of a (directed) link between two nodes \cite{YaChZhGoJi10}. 
Most of these quality measures have natural extensions to weighted networks, in
which case expressions such as ``number of edges'' are replaced with ``total 
edge weight'' instead. 

In our paper, we propose a simple weighting scheme that converts a directed
graph into a weighted undirected graph. This model enables us to utilize
the richness and complexity of existing methods to find communities
in undirected graphs. In particular, various schemes have been 
developed to optimize those four types of connectivity metrics, see 
\cite{BlGuLaLe08,GlSe12,GoMoCl10,NeGi04,Ne06,WhGlDh13} among others.  

Our specific weighting scheme is based on extending the idea that, within
 ``good'' communities, information can be shared
within a community more easily than between communities.  As information can flow along edges,
short (directed) paths and (directed) cycles can be seen as important in the function of 
communities within networks.  For instance, a short path between two nodes indicates that information can travel
between them quickly---or can travel quickly from the source to the destination in the case of a 
directed path.  A short cycle indicates that information can travel quickly among a group of nodes and,
thus, is even more important in the indicating a community than a short path. The simplest
cycle is a path that follows an undirected edge and then returns, and the second simplest
cycle is a path that follows a triangle. Consequently, triangles are the basis of 
many community structures. They also arise because of homophily, the fact that friends of friends are likely to become
 friends. Directed networks, however, pose a  more complicated problem since there are 7 different types
 of triangles and their contribution  to the community structure can be interpreted differently.  Here, we focus
on the importance of reciprocated edges and directed 3-cycles. Our weighting schemes are designed to increase
the weight associated with edges involved in both of these scenarios when a given directed graph is
converted to a weighted undirected graph.

The specific contributions of our paper are:
\begin{itemize} 
	\item We introduce a scheme that uses information about directed triangles to improve community detection in directed networks.  
	\item Our scheme involves creating an undirected but \emph{weighted} version of the network, which allows 
	us to utilize the wealth of existing community detection schemes for undirected networks.
	\item We propose a new metric on the quality of directed communities based on the number of 
	edges contained in 3-cycles split across communities.
	\item We show up to a 50\% reduction in the number of cycles cut in a
	partitioning into communities compared to simply ignoring the direction
	of each edge without a meaningful change to existing community
	metrics.
\end{itemize}

%% file: Background.tex
\section{Background and Motivation.}

\subsection{Directed Triangles}

\begin{figure*}[t!]
\centering
\includegraphics{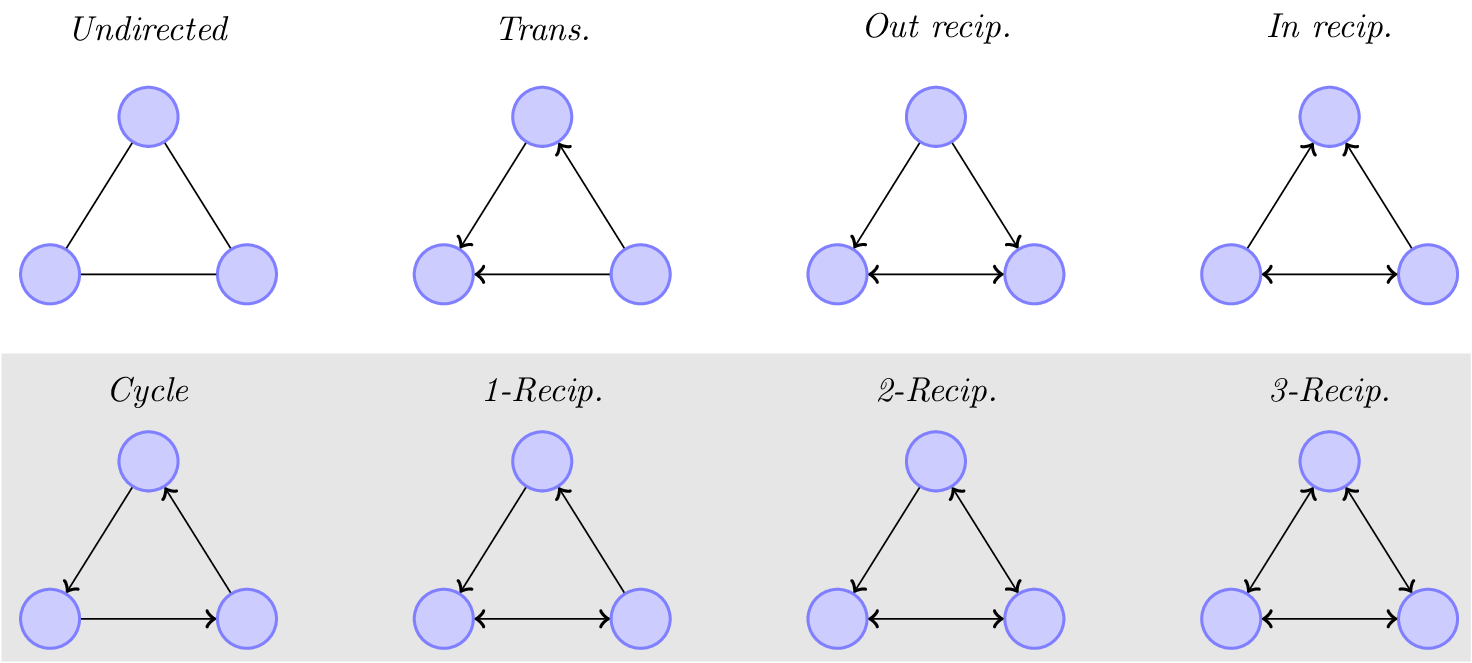}
\caption{An undirected triangle and the seven types of directed triangles.  The four triangles in the bottom row all contain a directed 3-cycle and form the basis of our weighting scheme to indicate community structure in directed networks; the three directed triangles in the top row only indicate partial information flow and we do not use them in our weightings. }
\label{fig:triangle_types}
\end{figure*}

In undirected networks, there is only one type of triangle.
Directed networks have seven triangle types (as observed
by, for instance,  \cite{SePiDuKo13}). We show the 
difference between these seven types in Figure~\ref{fig:triangle_types}.
Of these seven triangles, only a few are relevant for
community detection. Recall that the reason triangles are 
important in community detection in undirected networks is that the presence 
of a triangle indicates a mutually close relationship and 
ability to share information between three nodes.  
This is not the case for all types of directed triangles.  
Only triangles containing a 3-cycle (closed path of 
length three) enables information sharing among three nodes.  
The directed triangles 
which contain a 3-cycle are types in the bottom row of the
figure.

\subsection{Notation}
An  undirected network $G=(V,E)$ consists of a set of $|V|=n$ nodes and $|E|=m$ edges consisting 
of unordered pairs of nodes.  In a directed network, the edges are formed by ordered pairs of nodes.  
Let $d_i$ denote the degree of node $v_i$ in an undirected network.  In a directed network, each 
node $v_i$ has an in-degree, denoted $d_i^{{\rm in}}$, which is the number 
of edges that point into node $v_i$, an out-degree, $d_i^{{\rm out}}$, which is the number of edges pointing out of $v_i$, 
and a reciprocal degree, $d_i^{{\rm rec}}$, consisting of the number of reciprocal pairs of links in which 
node $v_i$ is involved.  The reciprocal edges do not contribute to the in- and out- degrees of node $v_i$.
 The adjacency matrix of a network $G$ is given by: 
$$A = (a_{ij}); \quad a_{ij}=\left\{\begin{array}{ll}
	1& \textnormal{ if } (v_i,v_j) \in E,\\
	0& \textnormal{ otherwise. }
	\end{array}\right .
	$$
If $G$ is undirected, $A$ will be symmetric and if $G$ is directed, $A$ will not be.  In the case of a
directed network,  let $A_s$ be the symmetric part of $A$ and $A_{ns}$ be the nonsymmetric part.  
Then, the unweighted, undirected representation of $G$ is given by the matrix $A_{ud} = A_s + A_{ns} + A_{ns}^T$. This is equivalent to simply
dropping the direction information on each edge in $G$.

A common quality measure for community assignment is {\em modularity} \cite{Ne03,Ne06}.  The 
modularity of a community assignment on an undirected network is given by
 $$ Q = \frac{1}{2m} \sum_{i,j}\left[A_{ij} - \frac{d_i d_j}{2m} \right] \delta(c_i, c_j)$$
 where  $c_i$ is the community membership of node $v_i$ and $\delta(c_i, c_j) = 1$ if $c_i = c_j$ 
 and 0 otherwise.  Modularity measures the difference between the observed density of edges within 
 communities and the expectation in a random network with the same degree distribution.  The concept 
 of modularity can be extended to directed networks through  {\em directed modularity} \cite{ArDuFeGo07}.  
 The directed modularity of a given community assignment is given by: 
 $$ Q_d = \frac{1}{m} \sum_{i,j}\left[A_{ij} - \frac{(d_i^{{\rm out}}+ d_i^{{\rm rec}})( d_j^{{\rm in}} + d_j^{{\rm rec}})}{m} \right] \delta(c_i, c_j).$$

A more recently developed measure of community quality is based on the probability of a node 
staying within a community during a ``Google-like'' random walk on the network.  Called {\em LinkRank}, 
it is defined by: $$Q_{LR} = \sum_{i,j} \left[L_{ij} - \pi_i \pi_j\right]\delta(c_i,c_j)$$ where $L_{ij} = \pi_iG_{ij}$ 
where $G = \alpha \bar{A}+(1-\alpha)(\frac{1}{n}){\bf 1}{\bf 1}^T$ and $\pi$ is the PageRank vector 
\cite{KiSoJe10}.  The LinkRank of a community measures the difference between the observed percentage 
of time a random walk stays inside communities and the amount of time expected in a random network.  
For more information on ``Google-like'' random walks and the PageRank vector, see \cite{LaMe06}.

\subsection{Related Work.}

\textbf{Triangle structure.}
The existence of triangles has been shown to be important in the formation of complex networks, 
especially those with an underlying community structure 
\cite{DuPiKoSe12,PrDoBrJo12,SeKoPi12,SePiDuKo13}.  Intuitively, a group of nodes that is highly 
connected is a better community than a group of nodes that is less connected, and
a clique (i.e., a group of  nodes in which every node is directly connected to every other node) contains the maximum number of connections possible and
is considered a good community under most metrics; likewise for near-cliques (i.e., the addition of a few edges would form a clique).  Both of these structures are characterized by the presence
of many triangles.  Additionally, triangle structure has been shown to be important in community detection 
in undirected networks \cite{BeHeLaPh11,PrDoBrJo12,SeArGo11}. For example
in \cite{PrDoBrJo12}, the authors define a ``good'' community to be a group of nodes that is 
dense in terms of triangles and  introduce a community quality metric called {\em Weighted 
Community Clustering} (WCC).  They experimentally show that a community with high WCC will be 
denser and contain more triangles than a community with high modularity and low WCC. The use of 3-node motifs (triangles and wedges) to identify communities of different types was 
introduced in \cite{SeArGo11}.  The authors introduce a generalized version of modularity 
which takes these motifs into account and a spectral algorithm for approximating its maximum.

%% Need to revise
\textbf{Weighting.}
It is well known that most community finding methods are heuristics and suffer from
many potential faults. For instance, communities detected by these schemes can overlook important 
network characteristics \cite{FoBa07}. Weighting schemes often enable simple algorithms
to overcome these faults.  In undirected networks, weighting edges based on the 
number of triangles in which the edge is involved has improved the quality of communities 
\cite{BeHeLaPh11}. Additionally, this weighting scheme enables the Clauset, Newman, and Moore algorithm to discover 
communities smaller than the resolution limit of modularity. Other weighting schemes have also been used successfully  
\cite{KhRaHa11}. In directed networks, the number of reciprocal (bidirectional) edges cut by 
community assignments has been used as a measure of the ``goodness'' of a community \cite{LiZhBa12},
and a weighting scheme based on the stationary distribution of a directed graph was also
shown to generalize the notation of conductance in a network~\cite{Chung2005-directed-laplacian}. 

\textbf{Algorithms.} The majority of the quality optimization algorithms for community detection have been developed for undirected 
networks, especially those which are efficient enough to be applied to datasets of large size 
\cite{BlGuLaLe08,GoMoCl10,NeGi04}.  However, optimizing a community quality measurement 
(e.g.~modularity) on the underlying unweighted, undirected network ignores important information about the direction of the links as we show via the next example.

%% file: Metric.tex
\section{The cycle cut ratio metric}
\label{sec:metric}

We now introduce a new metric to measure the quality of a 
directed community assignment:

\bgroup
\noindent \textbf{Definition} ({$k$-cycle cut ratio}) \itshape
The $k$-cycle edges of a graph are those that are contained 
within any directed length $k$-cycle. 
Given a partition of the vertices
of the network, the $k$-cycle cut ratio is the fraction of
$k$-cycle edges cut by the partition.
\egroup

The measure generalizes the number of reciprocal edges cut
in a directed network, which is equivalent to the the 
$2$-cycle cut ratio. Due to the importance of triangles
in the formation of network communities, we propose that
the $3$-cycle cut ratio is an important new metric 
to evaluate directed communities.

Let us demonstrate this idea through an example.
Consider the network in Figure 2.  Nodes 1-5 form a clique as do nodes 6-10 and 
11-13.  Node 15 sits between the cliques.  This node can both send and receive information to the first
clique (nodes 1-5), but can only send information to the second (nodes 6-10) and can only receive 
information from the third (nodes 11-13).  

Let community A be the community assignment 
where node 15 is grouped with nodes 1-5 and community B be where node 15 is grouped with nodes 6-10.  
Intuitively, node 15 should be grouped in a community with nodes 1 through 5 because those are the nodes 
that node 15 can both send information to and receive information from.

However, the quality of the community assignment measured under either undirected or directed modularity 
is the same whether node 15 is grouped with nodes 1 to 5 or with nodes 6 to 10.  The community quality 
measured by the LinkRank of the community assignments is higher for community A, somewhat supporting our hypothesis.  The values for these measures can be found in Table 1. This example shows 
the importance of measuring how many directed 3-cycles are cut in a community assignment.  When only the 
number of reciprocal edges cut is considered, the two community assignments are the same.  By minimizing
 the number of 3-cycles cut, it becomes clear that node 15 should be grouped with nodes 1-5.

We now show that giving more importance to edges involved in directed triangles that include $3$-cycles decreases 
the number of $3$-cycle cut ratio (and often the number of reciprocal edges which are cut) even when the networks
are partitioned using an undirected weighted graph partitioning scheme.

\begin{figure}[H]
\centering

\vspace{0pt}
\includegraphics[width=\linewidth]{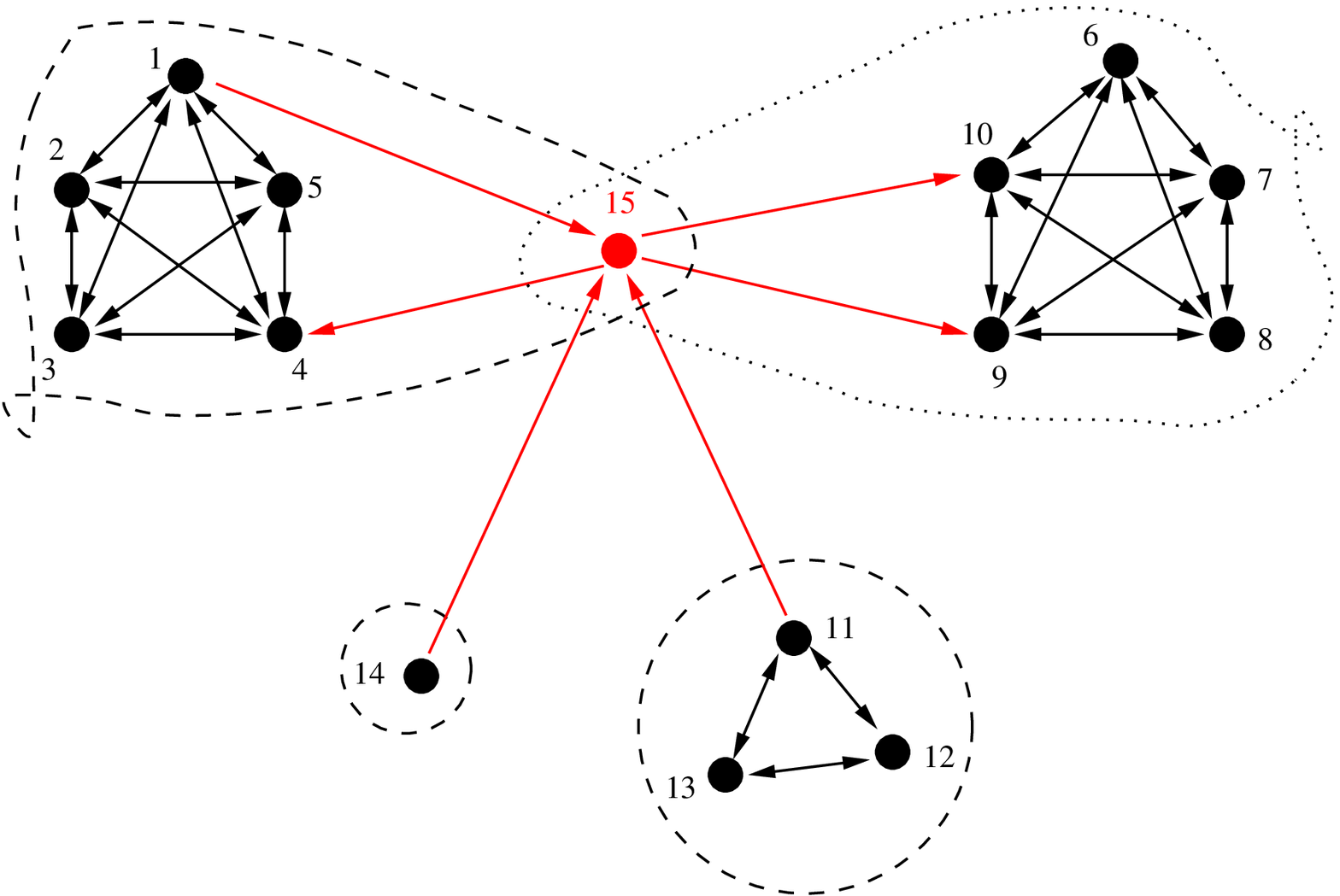}
\caption{It is not clear whether node 15 should be grouped in a community with nodes 1-5 or with nodes 6-10.}
\label{fig:small_exmaple}

\bigskip

\centering
\begin{tabularx}{\linewidth}{lXXX}
\toprule
 & $Q$ & $Q_d$ & $Q_{LR}$ \\
\midrule
 comm A & 0.4703 & 0.5318 & 0.4574 \\
comm B & 0.4703 & 0.5318 & 0.4465 \\
\bottomrule
\end{tabularx}
\captionof{table}{Comparison community quality metrics for two community assignments of the nodes in the network in Figure 2 when edges are unweighted.}
\label{tbl:small_example}
%\label{small_example}
\end{figure}

%% file: Methods.tex
\section{3-cycle weighting} 
\label{sec:methods}

Now that we have illustrated the importance of $3$-cycles in the
communities of a network, we wish to design a weighting scheme 
to turn a directed network into a weighted undirected network
with large weights on the edges in the $3$-cycles.

To calculate the weights for the 3-cycle weighting, we need to know the type of 3-cycles in which each edge participates.  
For each edge, we wish to know the following information:
\begin{itemize}
\item is it in a directed 3-cycle?
\item is it in a directed 3-cycle with one reciprocal edge?
\item is it in a directed 3-cycle with two reciprocal edges?
\item is it in a directed 3-cycle with three reciprocal edges?
\end{itemize}
Let $r_0, r_1, r_2, r_3$ be indicator vectors over the edges
of the graph denoting whether that edge is involved in any
of these cases. Once we have this information, we compute a 
weighting vector over the edges:
\[ w = \text{max}( 4 r_4, 3 r_2, 3 r_1, 2 r_0, 1 ). \]
Once we have this weight for each directed edge, we 
convert to an undirected weighted network by taking
the maximum weight of edge $(i,j)$ and $(j,i)$. 
(Any edge that does not exist has weight 0.)

A simple strategy to compute these vectors begins by building a list
of all directed cycles in the network. For each cycle, we then check
on the number of reciprocal edges, and then update indicator vectors
$r_0, r_1, r_2, r_3$ for each edge involved in the triangle. Our
implementation uses this strategy, however, we walk through the list
of directed 3-cycles algorithmically without writing out an explicit
list. This simplification greatly accelerates the computation as there
are often an incredibly large number of
directed cycles, and building an explicit in-memory list is
expensive. The strategy to walk the list implicitly starts
with a directed edge $(s,t)$, indexes the set of out-neighbors of $t$,
and then searches the set of in-neighbors of $s$ for any common vertex
with the out-neighbors of $t$. We then check the reciprocal status for
each edge in this cycle and update the appropriate vector.

%Once the indicator vectors are fully updated, the overlap between indicator vector $r_2$ and $r_3$ is
%zeroed out in $r_2$.  Then, the overlap between $r_2+r_3$ and $r_1$ is overlapped is zeroed out of $r_1$.
%This is to prevent doubly applying edge weights.  To apply the edge weights, first the matrix is symmetrized:
%$A_{ud} = A_s + A_{ns} + A_{ns}^T$.  Then the undirected network is weighted in the following manner:
%$A_{3cycle} = A_{ud} + r_0^T A_{ud}r_0 +r_1^T A_{ud}r_1 + r_2^T  A_{ud}r_2 + 2r_3^T  A_{ud} r_3$. 
%As every vector in a 3-cycle with one or more reciprocal edges will also be listed as in a directed 3-cycle, the
%above equation will produce a matrix with the appropriate edge weights.
%
%
%
%

%% file: Evaluation.tex
\section{Empirical Evaluation}

Our goal is to evaluate if our new 3-cycle based
weighting scheme yields improved communities of directed networks. Thus,
we examined the effects of edge-weighting on the directed modularity, LinkRank, the percentage of reciprocal
edges cut, and the percentage of edges contained in 3-cycles  cut during community partitioning on a number
of real world networks. The weightings we evaluate are:
\begin{itemize}
	\item {\em unweighted}: The undirected, unweighted network is used
	\[ A_{ud} = A_s + A_{ns} + A_{ns}^T. \]
	\item {\em reciprocal}: the underlying network is used, with edges that were reciprocal in the original 
network being given weight 2 and the remaining edges being given weight 1: 
\[ A_{r} = 2 A_s + A_{ns} + A_{ns}^T. \]
 \item \emph{3-cycle}: the 3-cycle weighting scheme we proposed in the last section.
\end{itemize}

We first review the method we use to identify communities in each network, then review the networks we study, and finally show the results of our weightings on each network.

\subsection{Community detection}

For
the task of extracting communities from an undirected, weighted network, we use METIS, a well established and understood
community detection method that is easy to adapt to edge-weights on a
network. 

The METIS software is a high-quality implementation of a multi-level
graph partitioning method~\cite{karypis1998-metis,AbKa06}. It constructs a
multi-level representation of an input graph by merging nodes and edges to form coarser representations, partitions the
coarsest graph, and then propagates and
refines the partitioning as it un-coarsens. It was originally designed to yield
balanced, minimum edge-cut-style partitions suitable for distributed
computing; yet it also produces useful sets for
clustering~\cite{Karypis2002-Cluto} and community detection. For
community detection, in particular, METIS is often used as a benchmark
or baseline method \cite{WaLoTaHo11}.

Each network was partitioned into 5, 10, 25, 50, and 100 communities using the three weighting schemes.

\subsection{Network data}

\begin{table*}
\centering
\caption{Basic network data for the 9 networks that we use in our empirical evaluation. We report the number of vertices $n$, number of directed edges $m$, the number of reciprocal edges, and the fraction of reciprocal edges. Then we also list the number of directed $3$-cycles with 0, 1, 2, and 3 reciprocal edges.} \vspace{1ex}
\begin{tabularx}{\linewidth}{llXXXXXXXX}
\toprule
 & Network & $n$ & $m$ & recip. & $r$ & 3-cycle & 1-recip.  & 2-recip. & 3-recip. \\ \addlinespace
\midrule 
1 & amazon0505 & 410K & 3,357K & 1,835K & 0.547 & 623 & 67k & 809k & 837k \\ \addlinespace
2 & Celegans & 297 & 2.3K & 394 & 0.168 & 72 & 179 & 148 & 16 \\ \addlinespace
3 & soc-Epinions1 & 76K & 509K & 206K & 0.405 & 7.7k & 84k & 328k & 160k \\ \addlinespace
4 & soc-Slashdot0902 & 82K & 948K & 810K & 0.854 & 92 & 10k & 77k & 406k \\ \addlinespace
5 & wb-cs-Stanford & 9.9K & 37K & 18K & 0.476 & 185 & 470 & 2197 & 6898 \\ \addlinespace
6 & web-BerkStan & 685K & 7601K & 1902K & 0.250 & 177 & 2185 & 12k & 72k \\ \addlinespace
7 & web-NotreDame & 326K & 1,470K & 759K & 0.517 & 9.5k & 41k & 107k & 6,780k \\ \addlinespace
8 & wiki-Vote & 8.3K & 104K & 5.9K & 0.057 & 6.8k & 18k & 15k & 2.1k \\ \addlinespace
9 & Wikipedia & 2118K & 28511K & 6131K & 0.215 & 553k & 2,529k & 3,929k & 1,091k \\ \addlinespace
\bottomrule
\end{tabularx}
\label{tbl:basicdata}
\end{table*}

We examined a total of 9 networks from a variety of real-world sources. Basic information on these
networks can be found in Table \ref{tbl:basicdata} including the number of nodes $n$, the number of edges
$m = \sum_{i=1}^n d_i^{in} + d_i^{rec}$, the number of reciprocal edges, given by $\sum_{i=1}^n d_i^{rec}$,
 and the reciprocity (percentage of edges which are reciprocal) $r = \left(\sum_{i=1}^n d_i^{rec}\right)/m$, and
 the number of 3-cycles of each type present in the network.
If any network had edge weights, we remove them before running our methods. All of the networks (other than the
Wikipedia network) can be found in the SNAP database \cite{SNAP}.  

The Wikipedia network is made up of the largest strongly connected component of the Wikipedia article-link graph, restricted to 
pages in categories containing at least 100 pages.  We used the Wikipedia article dump from 2011-09-01~\cite{wiki}. For each page, we also have category annotations from Wikipedia, these indicate topics within the encyclopedia that we use as a surrogate for communities.  We will make this data publicly available when this paper is published. In total, there are 17,364 categories. The average category contains $274$ pages and the median category contains $149$ pages. The largest category has 418k pages and includes all living people with pages on Wikipedia.

\subsection{Results}

\begin{figure*}[t!]
\centering
\includegraphics[width=0.95\textwidth]{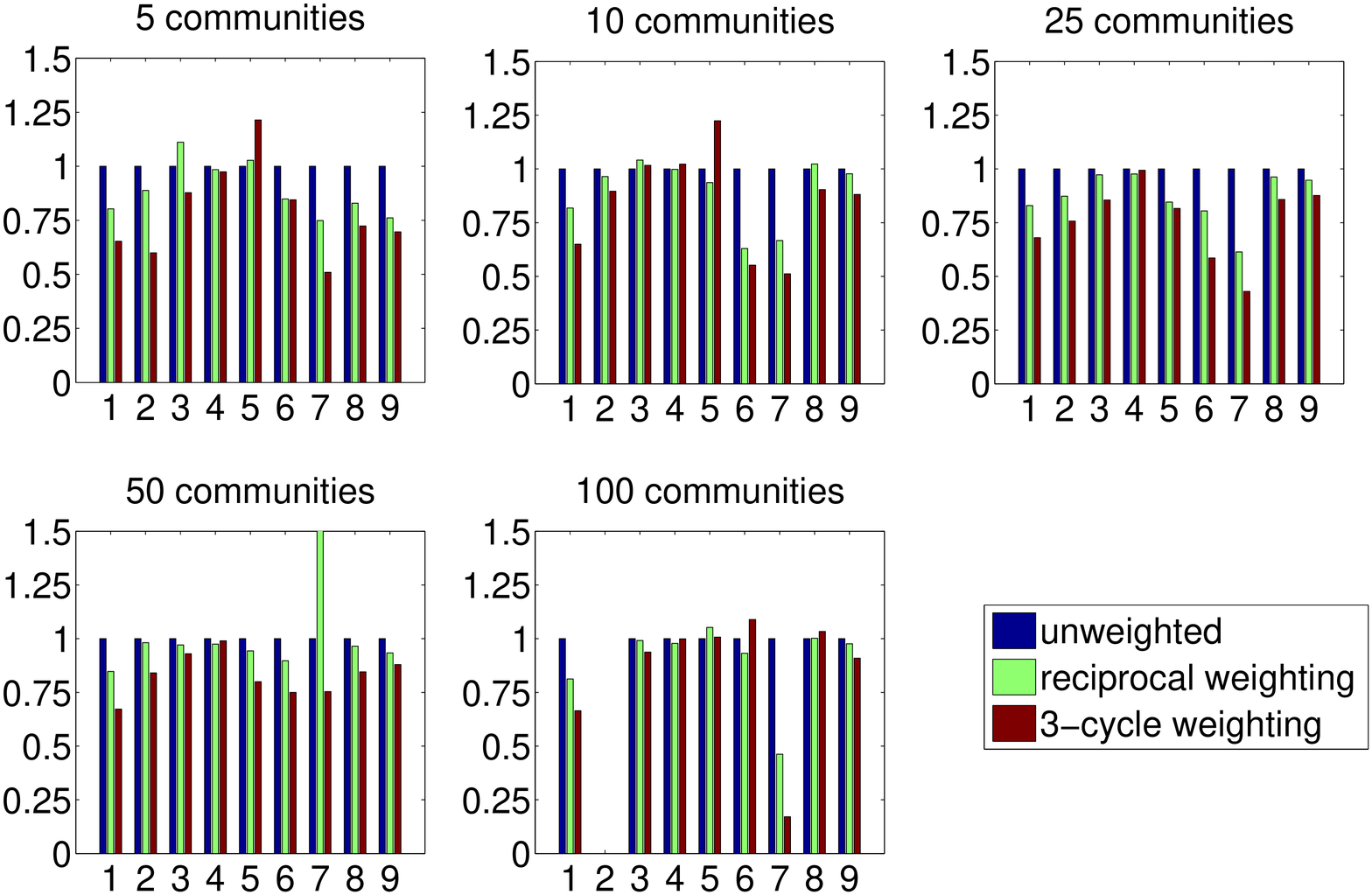}
\caption{The $3$-cycle cut ratio, relative to the unweighted case, for the nine networks considered (labeled according to their number in Table \ref{tbl:basicdata}) for various numbers of communities.  For network number 7 (the web-NotreDame network), the reciprocal weighting doubles $3$-cycle cut ratio, which is cut off in the plot. Overall, these results show that we are able to reduce the 3-cycle cut ratio by 10-50\%.}
\label{fig:bar_graph}
\end{figure*}

The directed modularity, LinkRank, number of reciprocal edges cut, and number of edges in 3-cycles cut 
in the networks under examination with the three weighting schemes are reported in Table
\ref{tbl:communities}.  The results are only reported for the number of communities that had the best overall 
directed modularity and LinkRank---see the table for the number of communities that resulted in the best.  
Table \ref{tbl:cut_3-cycles} then reports the percentage decrease in the number of 3-cycles cut under the reciprocal and 3-cycle based weighting schemes when compared to the unweighted partition; Figure~\ref{fig:bar_graph} displays the change in 3-cycle ratio visually.

\begin{table*}
\centering
\caption{The directed modularity, LinkRank, the fraction of reciprocal edges preserved, and fraction of 3-cycle edges preserved in the 9 networks under examination under the 3 weighting schemes described in Section \ref{sec:methods}.}
\begin{tabularx}{\linewidth}{lllXXXXX}
\toprule
& network & weighting scheme &  \# comm & $Q_d$ & $Q_{LR}$ &  rec. & 3-cycles \\
\midrule
\multirow{2}{*}{1} & \multirow{2}{*}{amazon0505} & unweighted & \multirow{2}{*}{50}  & 0.8768 & 0.7513 & 0.9382 & 0.9514 \\
& & reciprocal &  & 0.8740 & 0.7494 & 0.9515 & 0.9588 \\
& & 3-cycle &  & 0.8656 & 0.7496 & 0.9499 & 0.9673 \\
\midrule
\multirow{2}{*}{2} & \multirow{2}{*}{Celegans} & unweighted &  \multirow{2}{*}{5}  & 0.3799 & 0.2747 & 0.6193 & 0.6372 \\
& & reciprocal &  & 0.3835 & 0.2831 & 0.8020 & 0.6788 \\
& & 3-cycle &  & 0.3716 & 0.2967 & 0.8071 & 0.7823  \\
 \midrule
\multirow{2}{*}{3} & \multirow{2}{*}{soc-Epinions1} & unweighted &  \multirow{2}{*}{25} & 0.3982 & 0.3437 & 0.5406 & 0.5080 \\
& & reciprocal &  & 0.4069 & 0.3677 & 0.5913 & 0.5217 \\
& & 3-cycle &  & 0.3852 & 0.3614 & 0.5897 & 0.5791  \\
\midrule
\multirow{2}{*}{4} & \multirow{2}{*}{soc-Slashdot0902} & unweighted & \multirow{2}{*}{25} & 0.2943 & 0.3072 & 0.3444 & 0.3967 \\
& & reciprocal &  & 0.3033 & 0.3176 & 0.3633 & 0.4111 \\
& & 3-cycle &  & 0.2846 & 0.2942 & 0.3472 & 0.4005 \\
\midrule
\multirow{2}{*}{5} & \multirow{2}{*}{wb-cs-Stanford} & unweighted &  \multirow{2}{*}{50} & 0.8538 & 0.6482 & 0.9376 & 0.9187 \\
& & reciprocal &  & 0.8504 & 0.6541 & 0.9430 & 0.9233 \\
& & 3-cycle &  & 0.8516 & 0.6426 & 0.9283 & 0.9350 \\
\midrule
\multirow{2}{*}{6} & \multirow{2}{*}{web-BerkStan} & unweighted & \multirow{2}{*}{25} & 0.8918 & 0.7359 & 0.9969 & 0.9955 \\
& & reciprocal &  & 0.8912 & 0.7446 & 0.9985 & 0.9964 \\
& & 3-cycle &  & 0.8922 & 0.7362 & 0.9985 & 0.9974 \\
\midrule
\multirow{2}{*}{7} & \multirow{2}{*}{web-NotreDame} & unweighted & \multirow{2}{*}{100} & 0.9022 & 0.5443 & 0.9816 & 0.9811 \\
& & reciprocal &  & 0.9125 & 0.5440 & 0.9923 & 0.9913 \\
& & 3-cycle &  & 0.9128 & 0.5438 & 0.9966 & 0.9968 \\
\midrule
\multirow{2}{*}{8} & \multirow{2}{*}{wiki-Vote} & unweighted & \multirow{2}{*}{10} & 0.3509 & 0.2462 & 0.6156 & 0.4969 \\
& & reciprocal &  & 0.3579 & 0.2641 & 0.5975 & 0.4856 \\
& & 3-cycle &  & 0.3531 & 0.2659 & 0.6628 & 0.5454 \\
\midrule
\multirow{2}{*}{9} & \multirow{2}{*}{Wikipedia} & unweighted & \multirow{2}{*}{25} & 0.6065 & 0.4515 & 0.7979 & 0.7873 \\
& & reciprocal &  & 0.6090 & 0.4566 & 0.8259 & 0.7994 \\
& & 3-cycle &  & 0.5991 & 0.4593 & 0.8219 & 0.8146 \\
\bottomrule
\end{tabularx}
\label{tbl:communities}
\end{table*}

\begin{table*}
\centering
\caption{The percentage decrease in the number of 3-cycles cut under reciprocal and 3-cycle weighting as compared to the unweighted case for the community partition with the highest LinkRank. The results show that our weighting scheme can be highly effective in networks such as web-NotreDame.}
\begin{tabularx}{0.9\linewidth}{lXXXX}
\toprule
& Network & \# comm & reciprocal weighting & 3-cycle weighting \\
\midrule
1 & amazon0505 & 50 & 15.28\% & 32.77\% \\
2 & Celegans & 5 & 11.20\% & 40.00\% \\
3 & soc-Epinions1 & 25 & 2.78\% & 14.44\% \\
4 & soc-Slashdot0902 & 25 & 2.37\% & 0.64\% \\
5 & wb-cs-Stanford & 50 & 5.69\% & 20.06\% \\
6 & web-BerkStan & 25 & 19.55\% & 41.50\% \\
7 & web-NotreDame & 100 & 53.78\% & 82.91\% \\
8 & wiki-Vote & 10 & -2.25\% & 9.63\% \\
9 & Wikipedia & 25 & 5.24\% & 12.34\% \\
\bottomrule
\end{tabularx}
\label{tbl:cut_3-cycles}
\end{table*}

The first aspect of our data that we wish to highlight is that the weight schemes do not appreciably change the directed modularity or LinkRank scores. These indicate that the partitions we produce through this scheme have not greatly reduced the quality of the communities by traditional community detection metrics. Although the directed modularity and LinkRank of the community partitions found by weighting edges by 3-cycles do tend to be lower than those found in the unweighted case, the measures only decrease slightly.  And in some cases they increase.  The largest drop in directed modularity between the two sets of communities occurs
in the soc-Epinions1 network, where the directed modularity decreases by 0.0130.  The largest drop in LinkRank
occurs in the soc-Slashdot0902 network, where LinkRank also drops by 0.0130.  Neither of these decreases nor increases is
especially significant.

The second aspect we wish to note is that, for each of the networks examined (with the exception of web-BerkStan), weighting the edges of the network based on participation 
in 3-cycles significantly decreases the $3$-cycle cut ratio.    The fraction of preserved edges increases 
by about 0.015  from the unweighted case for the amazon0505 and web-NotreDame networks to an increase of almost 0.15 for the
Celegans network.  For all the networks examined (with the exception of the wiki-Vote network), the reciprocal edge-based 
weighting scheme also increases the number of 3-cycle edges
which are preserved, although not as significantly as under the 3-cycle based weighting.  

We refine our understanding of this change in Figure~\ref{fig:bar_graph} and Table~\ref{tbl:cut_3-cycles}. the results presented in Figure \ref{fig:bar_graph} demonstrate that weighting edges based on participation in 3-cycles improves
the ratio of 3-cycles cut in almost all partitions.  The exceptions are the wb-cs-Stanford network when 5 and 10 communities are 
considered and the web-BerkStan and wiki-Vote networks when 100 communities are considered.  In all other cases, weighting
edges by participation in 3-cycles reduces the 3-cycle cut ratio, often very significantly. 

Table \ref{tbl:cut_3-cycles} presents the relative effects of the reciprocal and 3-cycle based weighting schemes on the 3-cycle cut 
ratio compared to the unweighted case. For all the networks considered, with the exception of wiki-Vote network, reciprocal weighting also increases the 3-cycle cut ratio, 
although often not nearly as significantly as 3-cycle weighting; and, for all of the networks considered, 3-cycle weighting improves the 3-cycle cut ratio, often 
by a very significant margin.  In the web-NotreDame network, the 3-cut cycle ratio is improved by over 80\% compared to the 
unweighted case.  The only network where the improvement is not significant is the web-BerkStan network, which shows an improvement
of only 0.64\%.  
 
The third, and final aspect, we wish to mention that in the majority of cases (except wb-cs-Stanford), weighting the
edges based on participation in 3-cycles also significantly reduces the number of reciprocal edges cut.  For the Celegans, 
web-BerkStan, web-NotreDame, and wiki-Vote networks, the 
3-cycle based weighting scheme results in the greatest number of reciprocal edges being preserved.  For the amazon0505,
soc-Epinions1, soc-Slashdot0902, and Wikipedia networks, the number of reciprocal edges preserved increases from
the unweighted case but is lower than under the reciprocal edge-based weighting scheme.  The wb-cs-Stanford
network is the only network where the number of reciprocal edges preserved decreases from the unweighted case to the
case where weights are based on 3-cycles.

\subsection{Wikipedia categories}

Our final empirical evaluation consists of comparing the communities that result from our three weighting schemes to the categories in Wikipedia. We are primarily concerned with the number of intra-category edges that were cut by each of the schemes. These are the edges that are within the ground-truth communities in Wikipedia, but are separated by our methods. We use the partition of Wikipedia into 25 communities because that had the highest directed modularity.  The results are shown in the following table, which lists the number of edges cut by each of the methods, the number of within-community edges cut, the fraction of total cut edges that are within a community, and the improvement in that fraction relative to the unweighted scheme.

\begin{tabularx}{\linewidth}{@{}XXXll@{}}
\toprule
	&	cut edges	&	cuts \rlap{within} &	ratio	& 		\\
\midrule	
unweighted	&	10030459	&	768484	&	0.0766	&	0\%	\\
reciprocal	&	9958426	&	730711	&	0.0734	&	4\%	\\
3-cycle	&	10249256	&	743340	&	0.0725	&	5\%	\\
\bottomrule
\end{tabularx}

Both weighting schemes reduce the number of within-category edges cut and improve the fraction of within-community edges cut. Although the 3-cycle scheme actually cuts more within community edges than the reciprocal scheme, it also makes many more cuts in general, and thus its ratio of within community edges cut is lower.  We see these results as evidence that these weighting schemes are effective at improving detection of real world communities.

%% file: Conclusions.tex
We have described a simple weighting scheme to improve the detection of communities in 
directed networks. We also described a new metric for directed communities,
the $3$-cycle cut ratio. When we use our weighting scheme to convert a directed
network into a weighted undirected network and apply a standard 
network partitioning tool, we find a substantial reduction in
the $3$-cycle cut ratio, without any appreciable change in the traditional community
detection metrics such as modularity. Due to the simplicity of this approach, and the 
property that it reduces to a weighted, undirected network that can be 
analyzed with any new method, we are optimistic that this scheme will be
used by others to study directed community detection.

There are a few ways to continue investigating this idea. First, it is known 
that real world community often overlap. Thus, it would likely be fruitful
to to study overlapping community detection on the weighted undirected
graphs from our method. Second, the current set of weights assigned 
to each edge was not optimized at all; we conjecture that it will
be possible to further improve upon our results by optimizing
these weights for specific types of graphs. This step, however, requires
care not to overtune the weights to a particular type of network.
 Third, it is possible that using 
easy-to-compute graph structures such as biconnected components, 
$k$-cores, and other well-studied features may enable faster 
community detection in light of these weights and the directed 
triangle structure.

% Overlapping
% k-cores
% Variations on weights
% 

%% file: directed_communities.bbl
\begin{thebibliography}{10}

\bibitem{AbKa06}
A.~Abou-Rjeili and G.~Karypis.
\newblock Multilevel algorithms for partitioning power-law graphs.
\newblock In {\em IEEE International Parallel \& Distributed Processing
  Symposium (IPDPS)}, pages 10--27, 2006.

\bibitem{ArDuFeGo07}
A.~Arenas, J.~Duch, A.~Fern\'andez, and S.~G\'omez.
\newblock Size reduction of complex networks preserving modularity.
\newblock {\em New Journal of Physics}, 9:176--190, 2007.

\bibitem{BeHeLaPh11}
J.~W. Berry, B.~Hendrickson, R.~A. LaViolette, and C.~A. Phillips.
\newblock Tolerating the community detection resolution limit with edge
  weighting.
\newblock {\em Phys. Rev. E}, 83(5):056119, May 2011.

\bibitem{BlGuLaLe08}
V.~D. Blondel, J.-L. Guillaume, R.~Lambiotte, and E.~Lefebvre.
\newblock Fast unfolding of communities in large networks.
\newblock {\em Journal of Statistical Mechanics: Theory and Experiment},
  2008(10):P10008, 2008.

\bibitem{BoLaMoChHw06}
S.~Boccaletti, V.~Latora, Y.~Moreno, M.~Chavez, and D.-U. Hwang.
\newblock Complex networks: Structure and dynamics.
\newblock {\em Physics Reports}, 424:175--308, 2006.

\bibitem{BrEr05}
U.~Brandes and T.~Erlebach, editors.
\newblock {\em Network Analysis: Methodological Foundations, Lecture Notes in
  Computer Science Vol. 3418}.
\newblock Springer, New York, 2005.

\bibitem{Chung2005-directed-laplacian}
F.~Chung.
\newblock Laplacians and the {Cheeger} inequality for directed graphs.
\newblock {\em Annals of Combinatorics}, 9(1):1--19, 2005.
\newblock 10.1007/s00026-005-0237-z.

\bibitem{DuPiKoSe12}
N.~Durak, A.~Pinar, T.~G. Kolda, and C.~Seshadhri.
\newblock Degree relations of triangles in real-world networks and graph
  models.
\newblock In {\em CIKM '12}, CIKM '12, pages 1712--1716. ACM, 2012.

\bibitem{Es11}
E.~Estrada.
\newblock {\em The Structure of Complex Networks}.
\newblock Oxford University Press, 2011.

\bibitem{Fo10}
S.~Fortunato.
\newblock Community detection in graphs.
\newblock {\em Physics Reports}, 486(3):75--174, 2010.

\bibitem{FoBa07}
S.~Fortunato and M.~Barth\'elemy.
\newblock Resolution limit in community detection.
\newblock {\em Proc. Natl. Acad. Sci.}, 104 (1):36--41, 2007.

\bibitem{GlSe12}
D.~F. Gleich and C.~Seshadhri.
\newblock Vertex neighborhoods, low conductance cuts, and good seeds for local
  community methods.
\newblock In {\em KDD'12}, pages 597--605, 2012.

\bibitem{GoMoCl10}
B.~H. Good, Y.~A. de~Montjoye, and A.~Clauset.
\newblock The performance of modularity maximization in practical contexts.
\newblock {\em Phys. Rev. E}, 81:046106, 2010.

\bibitem{Karypis2002-Cluto}
G.~Karypis.
\newblock {CLUTO} -- a clustering toolkit.
\newblock Technical report, University of Minnesota, Department of Computer
  Science, 2002.

\bibitem{karypis1998-metis}
G.~Karypis and V.~Kumar.
\newblock A fast and high quality multilevel scheme for partitioning irregular
  graphs.
\newblock {\em SIAM J. Sci. Comput.}, 20(1):369--392, 1998.

\bibitem{KhRaHa11}
A.~Khadivi, A.~A. Rad, and M.~Hasler.
\newblock Network community-detection enhancement by proper weighting.
\newblock {\em Phys. Rev. E}, 83:046104, 2011.

\bibitem{KiSoJe10}
Y.~Kim, S.~W. Son, and H.~Jeong.
\newblock Finding communities in directed networks.
\newblock {\em Phys. Rev. E}, 81:016103, 2010.

\bibitem{LaMe06}
A.~N. Langville and C.~D. Meyer.
\newblock {\em Google's PageRank and Beyond: The Science of Search Engine
  Rankings}.
\newblock Princeton University Press, 2006.

\bibitem{LiZhBa12}
Y.~Li, Z.-L. Zhang, and J.~Bao.
\newblock Mutual or unrequited love: Identifying stable clusters in social
  networks with uni-and bi-directional links.
\newblock In {\em WAW'12: Algorithms and Models for the Web Graph}, pages
  113--125. Springer, 2012.

\bibitem{MaVa13}
F.~D. Malliaros and M.~Vazirgiannis.
\newblock Clustering and community detection in directed networks: A survey.
\newblock arXiv:1308.0971, Aug. 2013.

\bibitem{Ne03}
M.~E.~J. Newman.
\newblock Mixing patterns in networks.
\newblock {\em Phys. Rev. E}, 67:026126, 2003.

\bibitem{Ne06}
M.~E.~J. Newman.
\newblock Modularity and community structure in networks.
\newblock {\em Proceedings of the National Academy of Sciences},
  103(23):8577--8582, June 2006.

\bibitem{Ne10}
M.~E.~J. Newman.
\newblock {\em Networks: An Introduction}.
\newblock Cambridge University Press, Cambridge, UK, 2010.

\bibitem{NeGi04}
M.~E.~J. Newman and M.~Girvan.
\newblock Finding and evaluating community structure in networks.
\newblock {\em Phys. Rev. E}, 69:026113, 2004.

\bibitem{PrDoBrJo12}
A.~Prat-P\'erez, D.~Dominguez-Sal, J.~M. Brunat, and J.~L. Larriba-Pey.
\newblock Shaping communities out of triangles.
\newblock In {\em CIKM'12}, pages 1677--1681, 2012.

\bibitem{SeArGo11}
B.~Serrour, A.~Arenas, and G.~S.
\newblock Detecting communities of triangles in complex networks using spectral
  optimization.
\newblock {\em Computer Communications}, 34:629--634, 2011.

\bibitem{SeKoPi12}
C.~Seshadhri, T.~G. Kolda, and A.~Pinar.
\newblock Community structure and scale-free collections of {Erd\H{o}s-R\'enyi}
  graphs.
\newblock {\em Physical Review E}, 85(5):056109, May 2012.

\bibitem{SePiDuKo13}
C.~Seshadhri, A.~Pinar, N.~Durak, and T.~G. Kolda.
\newblock Directed closure measures for networks with reciprocity.
\newblock arXiv:1302.6220, Feb. 2013.
\newblock revised Sept. 2013.

\bibitem{SNAP}
SNAP.
\newblock Stanford network analysis project.

\bibitem{wiki}
Various.
\newblock {Wikipedia XML} database dump from {September} 1, 2011.
\newblock Accessed from \url{http://dumps.wikimedia.org/enwiki/}, September
  2011.

\bibitem{WaLoTaHo11}
L.~Wang, T.~Lou, J.~Tang, and J.~E. Hopcroft.
\newblock Detecting community kernels in large social networks.
\newblock Technical report, Cornell University, Tsinghua University, 2011.

\bibitem{WhGlDh13}
J.~J. Whang, D.~F. Gleich, and I.~S. Dhillion.
\newblock Overlapping community detection using seed set expansion.
\newblock In {\em CIKM'13}, 2013.

\bibitem{YaChZhGoJi10}
T.~Yang, Y.~Chi, S.~Zhu, Y.~Gong, and R.~Jin.
\newblock Directed network community detection: A popularity and productivity
  link model.
\newblock In {\em SIAM Data Mining'10}, 2010.

\end{thebibliography}
